# Distilling Accurate Descriptors from Multi-Source Experimental Data for Discovering Highly Active Perovskite OER Catalysts

Jingzhou Wang[1#], Huachao Xie[1#], Yuanqing Wang[1*], Runhai Ouyang[1,2*]

[1]Materials Genome Institute, Shanghai University, Shanghai 200444, China

[2]Zhejiang Laboratory, Hangzhou 311100, China

[#]The authors contributed equally.

*Corresponding authors: rouyang@shu.edu.cn; yuanqingwang@shu.edu.cn.

**Abstract**: Perovskite oxides are promising catalysts for oxygen evolution reaction (OER), yet the huge chemical space remains largely unexplored due to the lack of effective approaches. Here, we report the distilling of accurate descriptors from multi-source experimental data for accelerated catalysts discovery by using the new method SCMT-SISSO that overcomes the challenge of data inconsistency between different sources. While many previous descriptors for the catalytic activity were proposed based on respective small datasets, we obtained the new 2D descriptor ($d_B$, $n_B$) based on 13 experimental datasets collected from different publications and the SCMT-SISSO. Great universality and predictive accuracy, and the bulk-surface correspondence, of this descriptor have been demonstrated. With this descriptor, hundreds of unreported candidate perovskites with activity greater than the benchmark catalyst $Ba_{0.5}Sr_{0.5}Co_{0.8}Fe_{0.2}O_3$ were identified from a large chemical space. Our experimental validations on five candidates confirmed the three highly active new perovskite catalysts $SrCo_{0.6}Ni_{0.4}O_3$, $Rb_{0.1}Sr_{0.9}Co_{0.7}Fe_{0.3}O_3$, and $Cs_{0.1}Sr_{0.9}Co_{0.4}Fe_{0.6}O_3$.

## 1. INTRODUCTION

Perovskite oxides are promising catalysts for improving the kinetics of oxygen evolution reaction (OER) in the energy conversion of electrochemical water splitting[1-3]. Though highly active perovskites were reported recently, the huge chemical space of perovskite oxides remains largely unexplored due to the lack of effective approaches and the complexity of OER[1]. This reaction takes place mainly via the adsorbate evolution mechanism (AEM)[4] or lattice-oxygen participation mechanism (LOM)[5,6], with concerted or non-concerted proton-electron transfer[7]. However, the process to determine the OER activity and corresponding mechanism on a new catalyst by experiments and/or computations can be expensive and time demanding, which hinders efficient catalysts discovery. Alternatively, efforts have being done to identify physical quantities (termed descriptors) that directly correlate with the catalytic activity to guide rapid catalysts development[8].

Many descriptors for the OER activity of perovskite catalysts have been proposed based on respective datasets from experiments and/or computations, for instance, the number of d-electrons of the B-site metals [9], binding energy of oxygen atom to the surface[4], difference of adsorption free energy between oxygen and hydroxyl[10], $e_g$ orbital filling[2], p-band center of lattice oxygen[11], charge-transfer energy[7], concentration of oxygen vacancy[12], and the ratio between the octahedral factor and the tolerance factor[13]. On close inspection, one finds that all of these descriptors were proposed based on relatively small datasets, e.g. around ten samples. It is yet to be explored how generalizable they are, and it would be significantly important if a descriptor that is predictive within the huge chemical space of perovskite oxides can be identified.

A large amount of experimental activity data of OER on perovskites in literature from the past decades offers us new opportunities to distill more predictive and universal descriptors with advanced data-driven methods. However, the challenge is the data from different publications are often inconsistent due to their different experimental techniques and conditions. For example, with the same electrolyte of 0.1 M KOH and potential of 1.6 V vs. RHE, the same perovskites of $LaMO_3$ (M=Cr, Mn, Fe) presented very different activities between Ref.[14] and Ref.[7]. Hong et al. compiled a large experimental dataset from several publications for statistical evaluation of descriptors for the OER activity[15]. They handled the inconsistency by data standardization, and identified important parameters such as number of d-electrons, charge-transfer energy, $e_g$ orbital filling, and structural factors. Yet, an accurate relationship between the activity and the physical parameters remains missing, which requires stronger methods to deal with the data inconsistency.

Multi-task learning (MTL) was found effective in finding a common descriptor for several related properties (tasks)[16-18]. For example, $\mathbf{y}^t = \beta_1^t \boldsymbol{x_1} + \beta_2^t \boldsymbol{x_2}$ is a two-term (2D) linear model of MTL, where $\mathbf{y}^t$ is the $t$th property, $\{\beta_j^t\}$ are the coefficients for the $t$th task, and $\{\boldsymbol{x_j}\}$ are the common variables shared by all the tasks. The key idea of MTL is that all the tasks share the same descriptor ($\{\boldsymbol{x_j}\}$), but can be with different $\{\beta_j^t\}$ to account for the inconsistency and heterogeneity between tasks. Considering the abovementioned problem of inconsistency between multi-source data, if each dataset of the OER activity from different sources is treated as a task, then MTL promises to identify common descriptors for all these datasets, with the inconsistency addressed by the flexibility of the coefficients.

In this work, we first modified the traditional MTL by constraining the sign of the coefficients between different tasks to be the same to improve the model interpretability. The reason is that when the sign of coefficients between different tasks are different, e.g. $\beta_1^1 < 0 \text{ and } \beta_1^2 > 0$ in the abovementioned MTL model, it means the dependency of the OER activity on the same variable $\boldsymbol{x}_1$ is opposite between the two tasks, leading to the confusion of how to vary the $\boldsymbol{x}_1$ to increase the $\mathbf{y}$. We implemented the sign-constrained multi-task learning within the framework of Sure Independence Screening and Sparsifying Operator (SISSO)[19], termed SCMT-SISSO. Then, with the SCMT-SISSO and multi-source experimental data, a new descriptor with great accuracy and universality was identified. Density functional theory (DFT) calculations were performed to understand the descriptor, and high-throughput screening was done by applying this descriptor for the discovery of highly active perovskite catalysts, followed with subsequent experimental validations.

## 2. RESULTS AND DISCUSSION

We collected 13 available experimental datasets, including 8 in overpotential ($\eta$(V)) and 5 in the logarithm of current density ($\log(i/(\mathrm{mA/cm^2}))$), totally 182 data, of the OER activity on perovskite catalysts from publications in 1980 to 2020, as shown in Figure 1(a). The source and size of each dataset are listed in Table S1. All the $\eta$ data were multiplied by the constant -10 in our machine learning so that the values of the target $y$ being in the form of either $-10\eta$ or $\log(i)$ has the same range, and the greater the $y$, the higher the activity. A number of important and easy-to-obtain physical parameters of the A/B-site metal ions of perovskite were considered as input variables for model building, including the oxidation state ($n_B$), d-electron number ($d_B$), Shannon ionic radius ($r_A, r_B$), Pauling electronegativity ($\chi_A, \chi_B$), bond dissociation energy of metal-oxygen diatomic molecule ($\varepsilon_B$), and the octahedral factor ($\mu$) and tolerance factor ($t$) of the perovskite structure. Other parameters such as the O-2p band center[11], charge transfer energy (CTE)[7], and $e_g$ orbital filling[2] were not considered because of their high cost to obtain from either experiments or computations.

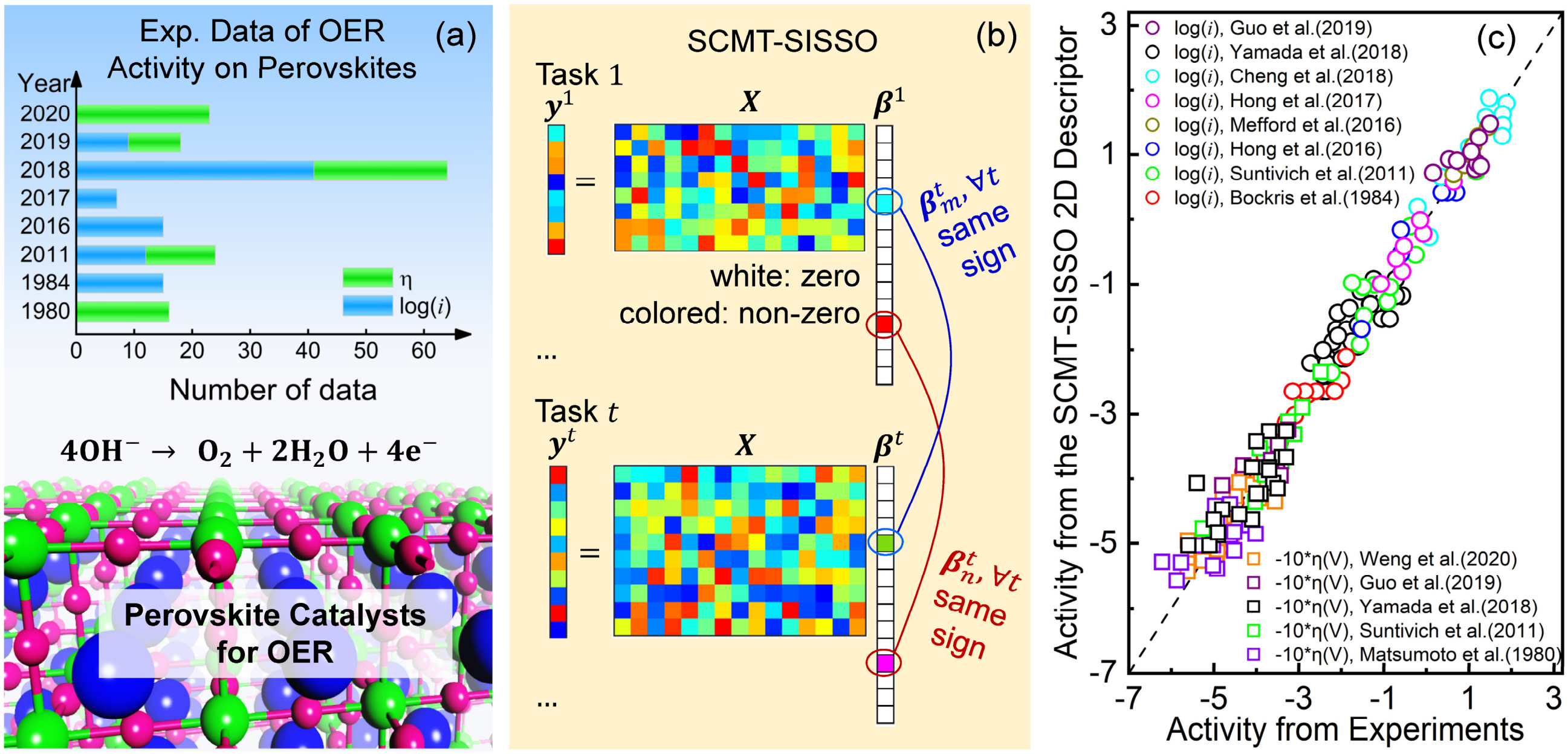

Figure 1. Descriptor identification for the OER activity of perovskite catalysts. (a) Distribution of the available experimental OER activity data on perovskite catalysts, published in the years from 1980 to 2020. (b) The idea of sign-constrained multi-task learning. The $i$th coefficients $\beta_i^t$ in all $t$ have the same sign. (c) Comparison of the activity data between the identified 2D descriptor ($d_B$, $n_B$) and the experiments. The colors denote the source of the datasets.

Usual symbolic regression with SISSO on the aggregated 13 datasets led to poor models for a series of complexity (coefficient of determination $R^2 \leq 0.10$ in training (Figure S1). After data standardization by adopting a previous scheme[15], the highest $R^2$ reaches 0.69 (Figure S2). To further improve the accuracy, we designed the following SCMT-SISSO, which is the traditional multi-task learning plus a sign-constraint, implemented within the framework of SISSO[18,19], to efficiently deal with the problem of data inconsistency (see also Figure 1(b)):

$$\widehat{\boldsymbol{B}} = \underset{\boldsymbol{B}}{\operatorname{argmin}} \sum_{t=1}^{T} \frac{1}{N^t} \|\boldsymbol{y}^t - \boldsymbol{X}\boldsymbol{\beta}^t\|_2^2 + \lambda\|\boldsymbol{B}\|_0 \tag{1}$$

$$\text{subject to: } \operatorname{sgn}\left(\beta_i^t \cdot \beta_i^{t'}\right) \geq 0 \quad \forall\, t, t'.$$

Here, the $\operatorname{sgn}$ is the sign function, $\beta_i^t$ is the $i$th element of the coefficient vector $\boldsymbol{\beta}^t$ for the task $t$, $B = \{\boldsymbol{\beta}^1, \ldots, \boldsymbol{\beta}^T\}$, $\boldsymbol{X}$ is the feature matrix that each column is a feature, $N^t$ is the number of samples in the task $t$, and the $\|\boldsymbol{B}\|_0$ is defined as the number of rows that have at least one nonzero element. The sign-constraint leads to the same sign for the $i$th coefficients $\beta_i^t$ of all the tasks. Equation (1) was solved by using the coordinate-descent algorithm[20]. By treating each of the 13 datasets as a task, the 13-tasks learning with the SCMT-SISSO were performed. As shown in Figure S3(a), the SCMT-SISSO models have much improved training accuracy (e.g. the obtained equation (2) shown below has the $R^2$ being 0.97 in training). Comparing to those from MT-SISSO (without sign-constraint), the SCMT-SISSO models have equivalent or slightly reduced accuracies but with improved interpretability.

A number of SCMT-SISSO models with increasing complexities were trained, and their prediction performance were evaluated by performing the leave-p-out cross validation (LPOCV), as shown in Figure S3. Specifically, in the LPOCV, we first randomly selected one data point from each of the 13 tasks, forming a test set containing 13 data in total. Then, with the remaining data, a SCMT-SISSO model was created to calculate the prediction errors on the test set. Such experiment was repeated 100 times, yielding the root-mean-square error (RMSE) over all the obtained 1300 prediction errors. Of all the considered models, the one with the lowest LPOCV RMSE (highest prediction accuracy) has the simple functional form (at dimension 2 & feature complexity 0):

$$\mathrm{y} = \beta_1^t d_B + \beta_2^t n_B + \beta_0^t, \tag{2}$$

where the $\mathrm{y}$ represents the OER activity, the $t$ from 1 to 13 denotes the 13 tasks, and the sign of coefficients are $\beta_1^t \geq 0$, $\beta_2^t \geq 0$, and $\beta_0^t < 0$ for all the tasks (Table S1).

The descriptor ($d_B$, $n_B$) is very stable against data removal/addition, i.e. it was identified 100 times in the 100 experiments of the LPOCV, indicating strong correlation between the descriptor and the OER activity. It is also accurate in describing all the datasets, as indicated in Figure 1(c). The all-data training RMSE of equation (2) is 0.32 (averaged over the 13 tasks), which corresponds to 39 mV (averaged errors for $\eta$) and 0.27 (averaged errors for $\log(i)$). The prediction RMSE of equation (2) from LPOCV is 0.43 (averaged over the 13 tasks), which corresponds to 47 mV for $\eta$, 0.40 for $\log(i)$. These accuracies are acceptable given that the training-data noise exist even in a single dataset (from one publication), e.g. the experimental error bar of measured overpotential can be up to 50 mV[2,11]. In Figure S3(c) we show that if noises (random numbers in [-25 mV, +25 mV]) were added to the target data, the $(d_B,\ n_B)$ remains the most predictive descriptor at fixed complexity from SCMT-SISSO.

It is immediately known from equation (2) that the OER activity increases with both the $d_B$ and $n_B$, which is a robust trend found in the 13 experimental datasets. Since both $d_B$ and $n_B$ are dimensionless quantities, the coefficients have the dimension of OER activity. The intercept $\beta_0$ represents the activity when extrapolated to $d_B = 0$ and $n_B = 0$ (there is no such perovskite oxide with zero $d_B$ and $n_B$). The first and second terms mean the two separate contributions to the activity modulated by the $d_B$ and $n_B$, respectively. Although both $d_B$ and $n_B$ are among many of the previously considered parameters[15], discovery of the simple 2D descriptor $(d_B,\ n_B)$ is nontrivial as it naturally emerge as the best out of huge number of candidates (different functional forms and complexity) based on 13 experimental datasets.

We note that both the $d_B$ and $n_B$ were obtained from the nominal stoichiometry of the compounds, and thus the same definition will be used throughout this paper. In addition, since the coefficients in different tasks are different, one may take the coefficients from a task based on the consideration of data quality, experimental methods, or other reasons. Before we demonstrate the application of equation (2), more analysis is presented below to gain deeper understanding of the descriptor $(d_B,\ n_B)$.

Figure 2(a) shows the comparison of accuracy between three different descriptors over the 13 datasets. Clearly, the 2D descriptor $(d_B,\ n_B)$ has systematically lower errors than the $d_B$[9] on all the experimental datasets, especially the ones by Hong et al.[15], Suntivich et al.[2], and Weng et al.[13]. The improvement of $(d_B,\ n_B)$ over $d_B$ is not simply due to one more coefficient to fit the data, but both the $d_B$ and $n_B$ are essential for comprehensive description of the B-site 3d-metal ions, e.g. distinguishing the $Co^{3+}$ from $Ni^{4+}$. The 2D descriptor also has systematically lower errors than the $\mu/t$ on all the datasets (except the one by Weng et al.[13]), with the most significant improvement on the datasets by Suntivich et al.[2] and by Bockris et al.[9].

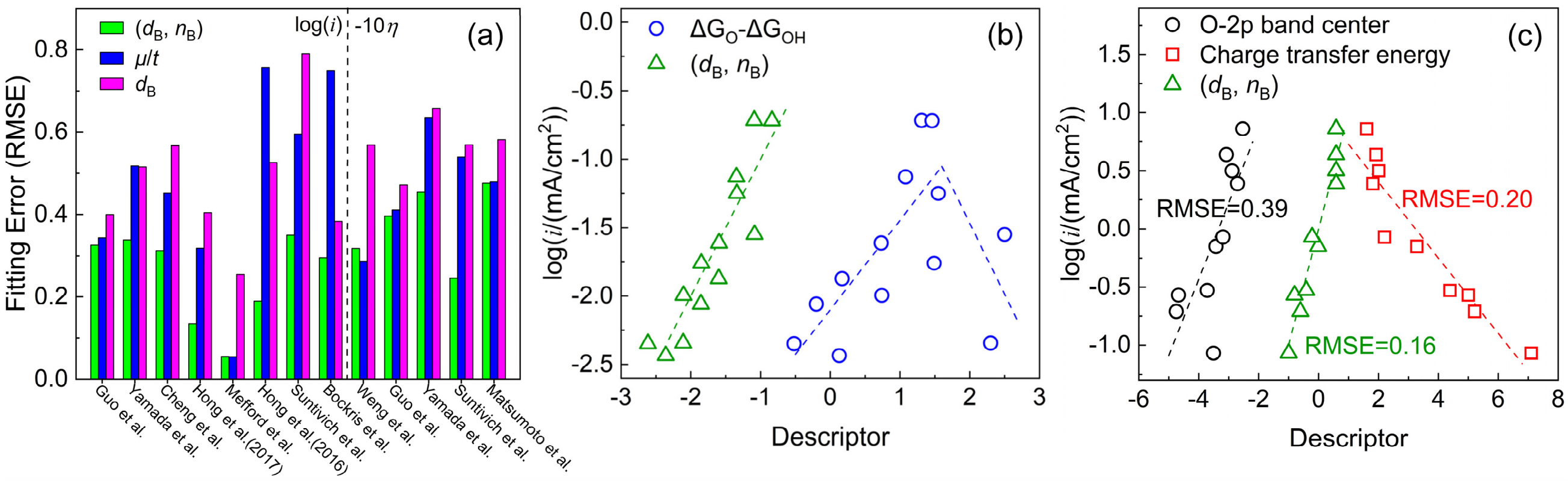

Figure 2. Comparison of accuracy between the $(d_B,\ n_B)$ and previous descriptors. The $(d_B,\ n_B)$ was converted to a single number by using the form of equation (2), with the coefficients determined by fitting to the dataset of interest. (a) The $(d_B,\ n_B)$, versus the $\mu/t$ and $d_B$ on the 13 datasets (8 for $\log(i)$ on the left of the dashed line, and 5 for $\eta$ on the right). (b) The $(d_B,\ n_B)$ versus the $\Delta G_O - \Delta G_{OH}$ on the experimental dataset from Ref.[14]. (c) The $(d_B,\ n_B)$ versus the O-2p center and the CTE on the experimental dataset from Ref.[7].

The $(d_B,\ n_B)$ was also compared to the descriptor $\Delta G_O - \Delta G_{OH}$[10]. Perfect volcano-like relationship of the computed over potential based on AEM versus the descriptor $\Delta G_O - \Delta G_{OH}$ was found[10]. However, if all those computed activity data were replaced by experimental values of corresponding perovskite catalysts taken from Ref.[14], they are no longer well described with the volcano-like relationship, as shown in Figure 2(b). This difficulty may be due to that the actual mechanisms on the catalysts are not always AEM as assumed in their calculations. In comparison, the $(d_B,\ n_B)$ shows much better accuracy on those experimental data, indicating that the $d_B$ and $n_B$ could be the key factors modulating the OER activity regardless of the mechanisms.

The $(d_B,\ n_B)$ was further compared to the descriptors of O-2p band center[11] and CTE[7]. It was found that high O-2p center favors the oxidation of lattice oxygen and the formation of oxygen vacancy, leading to high catalytic activity for OER[11]. Yet, the O-2p band center fails to correlate with the activity on insulator systems, and the CTE was then proposed as a better descriptor[7]. Figure 2(c) shows the comparison of both the CTE and the O-2p band center with the ($d_B$, $n_B$) on describing the experimental activity data from Ref.[7]. It is seen that ($d_B$, $n_B$) has the lowest error among the three. Importantly, the ($d_B$, $n_B$) are atomic parameters whose data can be easily obtained, which allows for rapid high-throughput catalysts screening, whereas the O-2p band center and CTE are of materials properties requiring experimental characterization or first-principles calculations.

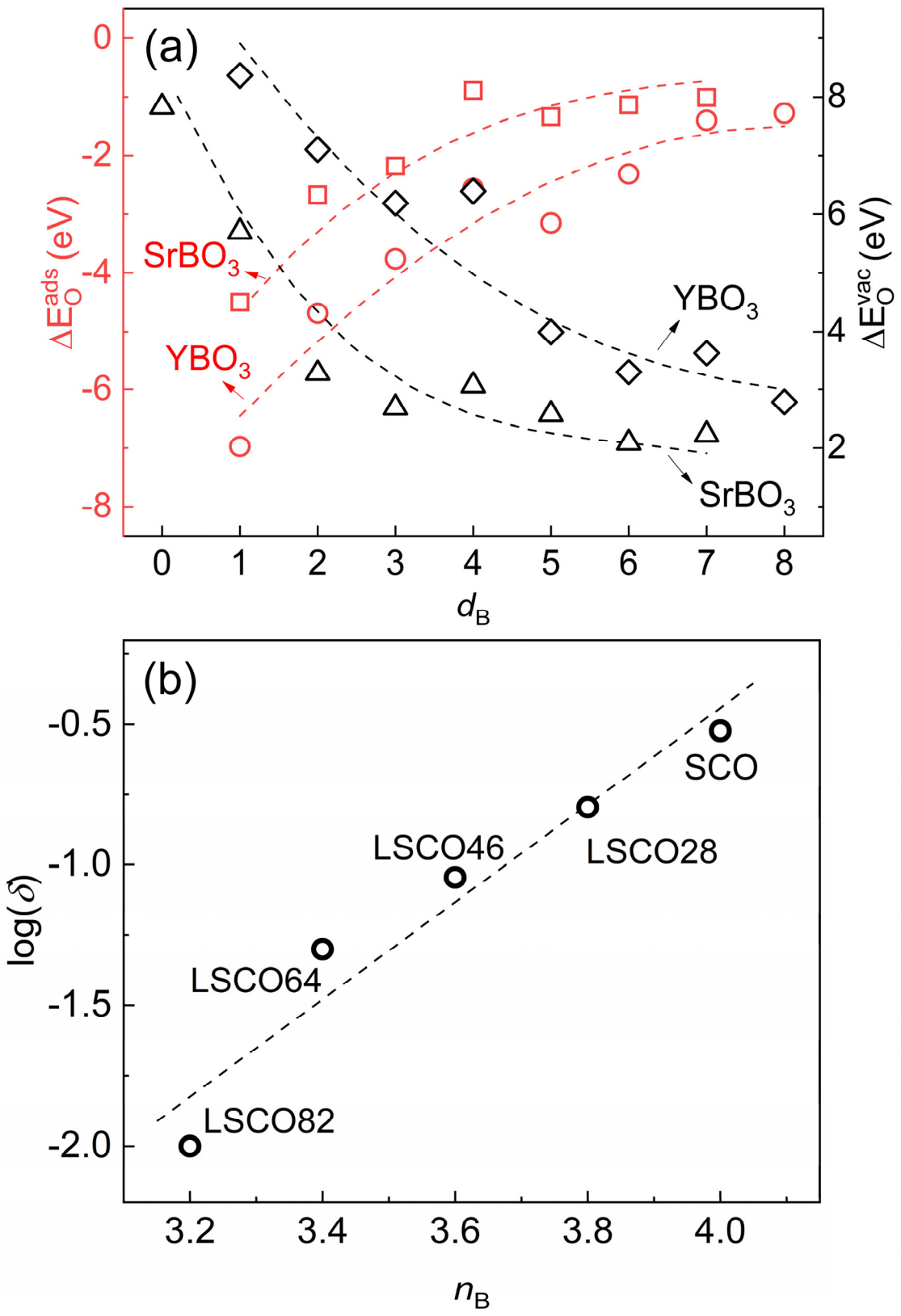

Figure 3. Correlation of the ($d_B$, $n_B$) with surface properties. (a) Correlation between the ($d_B$, $n_B$) and the calculated surface properties of O adsorption energy $\Delta E_O^{ads}$ and vacancy formation energy $\Delta E_O^{vac}$ on 3d-metal perovskites. The "B" in $SrBO_3$ and $YBO_3$ stands for the 3d-metals of Ti, V, Cr, Mn, Fe, Co, Ni, and Cu. Dashed lines are for guiding the eyes. (b) Correlation between the $n_B$ and the logarithm of the concentration of oxygen vacancy ($\delta$) on the $La_{(1-x)}Sr_xCoO_3$ systems. The $\delta$ are experimental data taken from Ref.[12]. $n_B = 6 - 3(1-x) - 2x$, with the $x$ of 0.2, 0.4, 0.6, 0.8, and 1.0 corresponding to LSCO82, LSCO64, LSCO46, LSCO28, and SCO, respectively.

Though the $d_B$ and $n_B$ are parameters from the bulk of perovskites, its strong correlation with the OER activity hints that it must correspond to key surface properties as catalysis happens on surfaces[21]. We performed DFT calculations on oxygen adsorption energy ($\Delta E_O^{ads}$) and oxygen vacancy formation energy ($\Delta E_O^{vac}$) on a series of B-site-3d-metals terminated (100) surfaces of single perovskites. Results in Figure 3(a) show that the $\Delta E_O^{ads}$ monotonically increases with $d_B$ at fixed $n_B$, indicating the weakened binding of the oxygen adsorbate with the 3d-metals from left to right of the periodic table of elements (PTE). This can be rationalized with the d-band center model that the early transition metals are more reactive than later ones[22,23]. In addition, $\Delta E_O^{ads}$ increases with $n_B$ going from $3^+$ to $4^+$ at fixed $d_B$, indicating weakened binding of the O adsorbate with a metal in higher oxidation states. Metal atoms with higher oxidation states tend to be more difficult to donate electrons for further oxidation[22]. Linear dependency of the $\Delta E_O^{ads}$ on each of the $d_B$ and $n_B$ was found in previous DFT calculations with different parameters [22]. It is worth mentioning that the correlation between the ($d_B$, $n_B$) and the $\Delta E_O^{ads}$ does not mean the experimental OER activity against the ($d_B$, $n_B$) or $\Delta E_O^{ads}$ has to comply with the "volcano-like" relationship[4], because the actual mechanism could not be AEM.

Given the crucial roles of oxygen vacancy on the catalysis, we also calculated the formation energy of surface oxygen vacancy ($\Delta E_O^{vac}$). Figure 3(a) shows the $\Delta E_O^{vac}$ decreases with $d_B$ at fixed $n_B$ because of the weakened oxygen-metal bonds as the B-site 3d-metal goes from left to right of the PTE. Similarly, the metal atoms with higher oxidation states (increased from $3^+$ to $4^+$) tend to have lower $\Delta E_O^{vac}$ because of weaker oxygen-metal bonds. The formation energy of bulk oxygen vacancy in perovskites with the $d_B$ and $n_B$ follow a similar trend from previous calculations[24]. Therefore, high $d_B$ and $n_B$ tend to favor the formation of oxygen vacancy both on surface and in bulk of perovskites.

Evidences of the correlation between the $n_B$ and the concentration of oxygen vacancy ($\delta$) in the perovskite systems of $La_{(1-x)}Sr_xCoO_3$ is shown in Figure 3(b), where the $\delta$ are consistent experimental data taken from the previous work[12]. For these materials, the ($d_B$, $n_B$) reduces to the single-parameter descriptor $n_B$, because $n_B + d_B = 9$. When the $x$ goes from 0 to 1, the $n_B$ changes from 3 to 4. Figure 3(b) shows the $n_B$ linearly increase with $\log(\delta)$. This is understandable considering that the $\delta$ was found linearly correlated with the OER current density ($i$)[12], and at the same time we found the $\log(i)$ linearly increase with the $n_B$ in the equation (2). The linear correlation between the $\log(\delta)$ and the ($d_B$, $n_B$) is further demonstrated in our experiments below.

These analyses explain why the activity generally increases with $d_B$ and $n_B$ in equation (2). Though the descriptor ($d_B$, $n_B$) itself does not imply any OER mechanism on a specific catalyst, increasing both $d_B$ and $n_B$ could avoid too strong O adsorption and facilitate the O-vacancy formation, favoring the OER no matter whether it proceeds via the AEM[4] or LOM[5,6]. However, one possible limitation of equation (2) is that it may fail to predict the OER activity for systems with high $d_B$ and $n_B$, e.g. $SrCuO_3$, $LaCuO_3$ and $SrTiO_3$, because of too weak of the O adsorption if the OER has to proceed via AEM[10].

Equation (2) was applied to screen for new active perovskite catalysts. We adopted the coefficients from the task whose training data (task 11 in Table S1) come from the previous work by Suntivich et al.[2] in which the highly active perovskite $Ba_{0.5}Sr_{0.5}Co_{0.8}Fe_{0.2}O_3$ (BSCF5582) was firstly discovered. Thus, our predicted OER activity refer to intrinsic activity to be consistent with the previous work[2]. The goal was to see if new perovskites with activity even higher than BSCF5582 can be identified. We considered the formula $A_xA'_{(1-x)}B_yB'_{(1-y)}O_3$, with $x, y \in [0, 1]$, and the elements of {Rb, Cs, Ca, Sr, Ba, La} for the A-site, the 3d metals {Ti, V, Cr, Mn, Fe, Co, Ni, Cu} for the B-site. By setting the step length for $x$ and $y$ to be 0.1, we obtained a chemical space containing 36660 possible compounds.

The screening was started with the filter of charge-balance, after which 26510 candidates survived. Then, after stability screening with the tolerance factor $0.8 < t < 1.1$ and the new tolerance factor $\tau \leq 4.18$[25], 16337 survived. Thirdly, after screening for higher activity (lower overpotential) than that of BSCF5582 with the equation (2), 880 candidates were obtained and are marked as green squares in Figure 4. In addition, we assigned the 1259 materials whose OER activities are predicted to be within $\Delta$ less active than BSCF5582 to be uncertain predictions and are marked as yellow triangles, where the $\Delta$ was taken as the training RMSE (24 mV) of the adopted task. All these “green” and “yellow” materials are listed in Table S2.

Figure 4 shows that nearly all of the “green” and “yellow” compounds contain one or more of the Fe, Ni and Co elements at the B-site, because of their high $d_B$ and $n_B$ according to equation (2). It serves as a useful “map” of OER activity for discovering new perovskites catalysts from the chemical space $A_xA'_{(1-x)}B_yB'_{(1-y)}O_3$. However, since the stability was only evaluated by tolerance factors, it cannot be excluded that other factors may hinder the actual synthesizability and long-term activity under operational conditions. For example, high oxidation states (> 2+) of Cu may be unstable, and the alkali metals at A-site can dissolve under electrochemical condition[26].

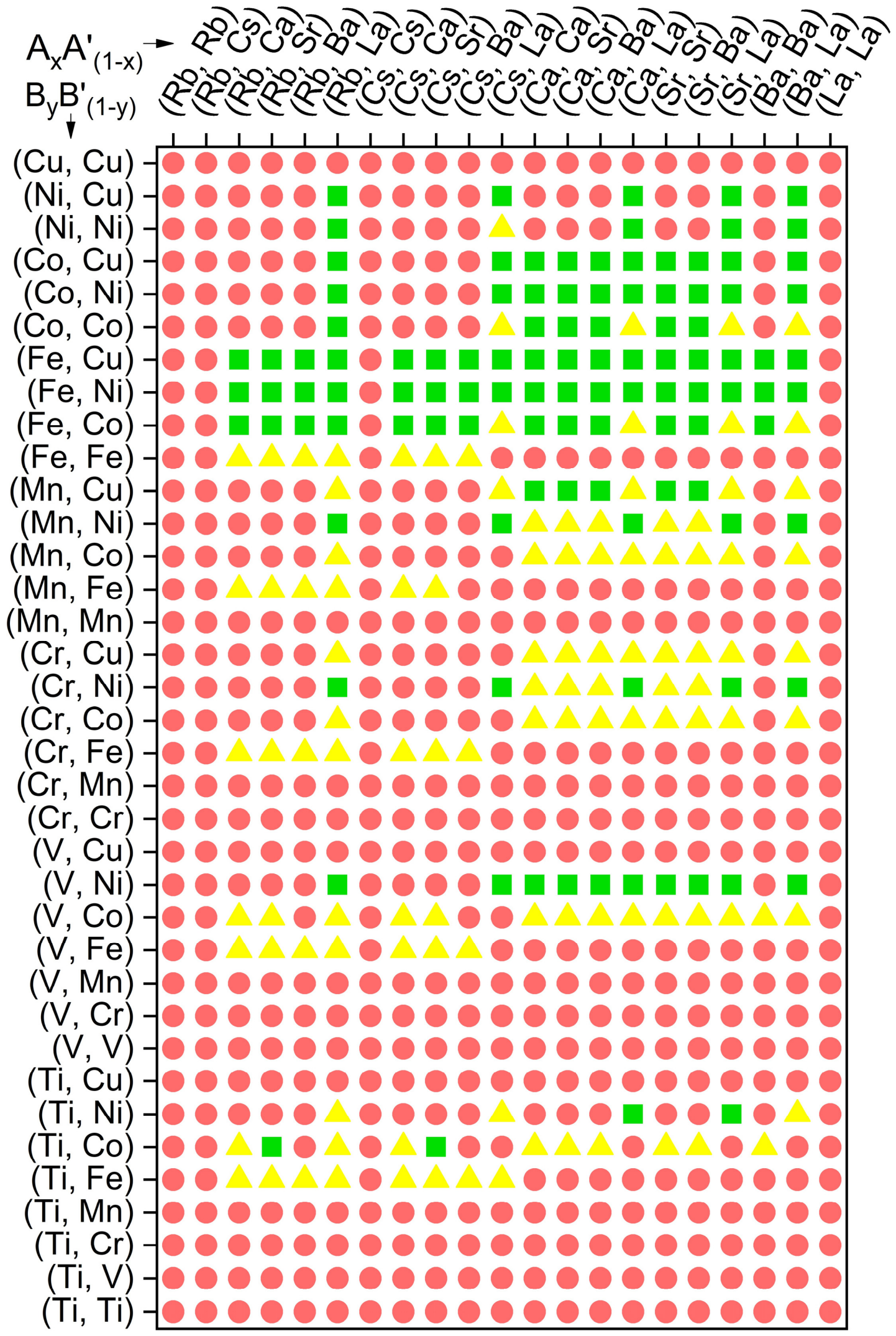

Figure 4. Screening for stable and highly active perovskite catalysts from the chemical space $A_xA'_{(1-x)}B_yB'_{(1-y)}O_3$. (a) Green: there is at least one point (x, y) that the equation (2) predicts the perovskite $A_xA'_{(1-x)}B_yB'_{(1-y)}O_3$ has lower overpotential than BSCF5582; Yellow: the overpotential predicted to be higher than BSCF5582 but lower than BSCF5582+Δ ($\Delta = 24$ mV, see the main text); Red: all the other $A_xA'_{(1-x)}B_yB'_{(1-y)}O_3$ that can not pass the screening.

One of the "green" materials, $Rb_{0.2}Sr_{0.8}Co_{0.4}Fe_{0.6}O_3$ (RSCF2846), was taken as an example to compare with the BSCF5582 for their electronic structures by performing DFT calculations with a 500-atoms SQS[27] supercell for each. The calculated projected density of states in Figure S4 shows that the RSCF2846 has higher O-2p band center (integrating over the occupied O-2p states) than the latter, being -2.29 and -2.35 eV, respectively. In addition, both RSCF2846 and BSCF5582 are metallic, and the RSCF2846 show smaller CTE. Given that high O-2p band center and small CTE are favorable for high OER activity[7,11], these electronic structures corroborate the prediction made by equation (2).

Experimental synthesis by using the sol-gel method was conducted on new materials mainly in the two series of $A_xA'_{(1-x)}Co_yFe_{(1-y)}$ and $SrB_yB'_{(1-y)}O_3$ to investigate their synthesizability and activity. They are 30 materials from the "green" list (except the CSCF2819 from the "yellow"), including 10 $Ba_xSr_{(1-x)}Co_yFe_{(1-y)}O_3$, 7 $Cs_xSr_{(1-x)}Co_yFe_{(1-y)}O_3$, 2 $Rb_xSr_{(1-x)}Co_yFe_{(1-y)}O_3$, 6 $SrB_yB'_{(1-y)}O_3$ (B,B' = V, Mn, Fe, Co, Ni, Cu) and 5 other Co, Ni, Cu based oxides. Our XRD results confirmed totally 26 perovskite structures, including 7 phase-pure cubic perovskites (all belong to $A_xA'_{(1-x)}Co_yFe_{(1-y)}$), 1 phase-pure hexagonal perovskite ($SrCoO_3$), 1 hexagonal perovskite but with minor un-identified phases ($SrCo_{0.6}Ni_{0.4}O_3$), and 17 others mainly in perovskite structures but with impurities or intermediate phases. The XRD patterns of three representative phase-pure $A_xA'_{(1-x)}Co_yFe_{(1-y)}$ materials: $Rb_{0.1}Sr_{0.9}Co_{0.7}Fe_{0.3}O_3$ (RSCF1973), $Cs_{0.1}Sr_{0.9}Co_{0.4}Fe_{0.6}O_3$ (CSCF1946), $Ba_{0.1}Sr_{0.9}Co_{0.9}Fe_{0.1}O_3$ (BSCF1991), and the $SrCo_{0.6}Ni_{0.4}O_3$ and $SrCoO_3$, as well as the confirmed phase-pure cubic perovskite BSCF5582 for benchmark are shown in Figure 5 (a), and all the others are provided in Figure S5. The interested materials in Figure 5 (a) were then experimentally evaluated for their catalytic activity for OER.

Figure 5(b) shows that the $SrCo_{0.6}Ni_{0.4}O_3$, RSCF1973 and CSCF1946 have lower overpotentials than the reference materials BSCF5582 and $RuO_2$. $SrCo_{0.6}Ni_{0.4}O_3$ has the highest activity, with the overpotential at 1 $mA \cdot cm^{-2}_{ECSA}$ to be 304 mV which is 108 mV lower than that of BSCF5582 (412 mV) and $RuO_2$ (412 mV). The OER activities of the tested perovskites follow the trend: $SrCo_{0.6}Ni_{0.4}O_3$ > RSCF1973 > CSCF1946 > BSCF5582 > $SrCoO_3$ > BSCF1991. Thus, 3 out of the 5 predicted intrinsic activity were confirmed to be higher than that of BSCF5582. Although the activity of $SrCoO_3$ and BSCF1991 disagree with the prediction, they are quite close to that of BSCF5582, i.e. within 20 mV at 1 $mA \cdot cm^{-2}_{ECSA}$. Note that the measured activities of these two benchmark catalysts reported in literature strongly depends on the preparation methods and test conditions (Table S3). Our activity data from $RuO_2$ and BSCF5582 fall in the range of previously reported values, though it is hard to make direct comparisons. In this work, the preparation methods and test conditions are fixed for all the materials, allowing for fair comparisons of the OER activity.

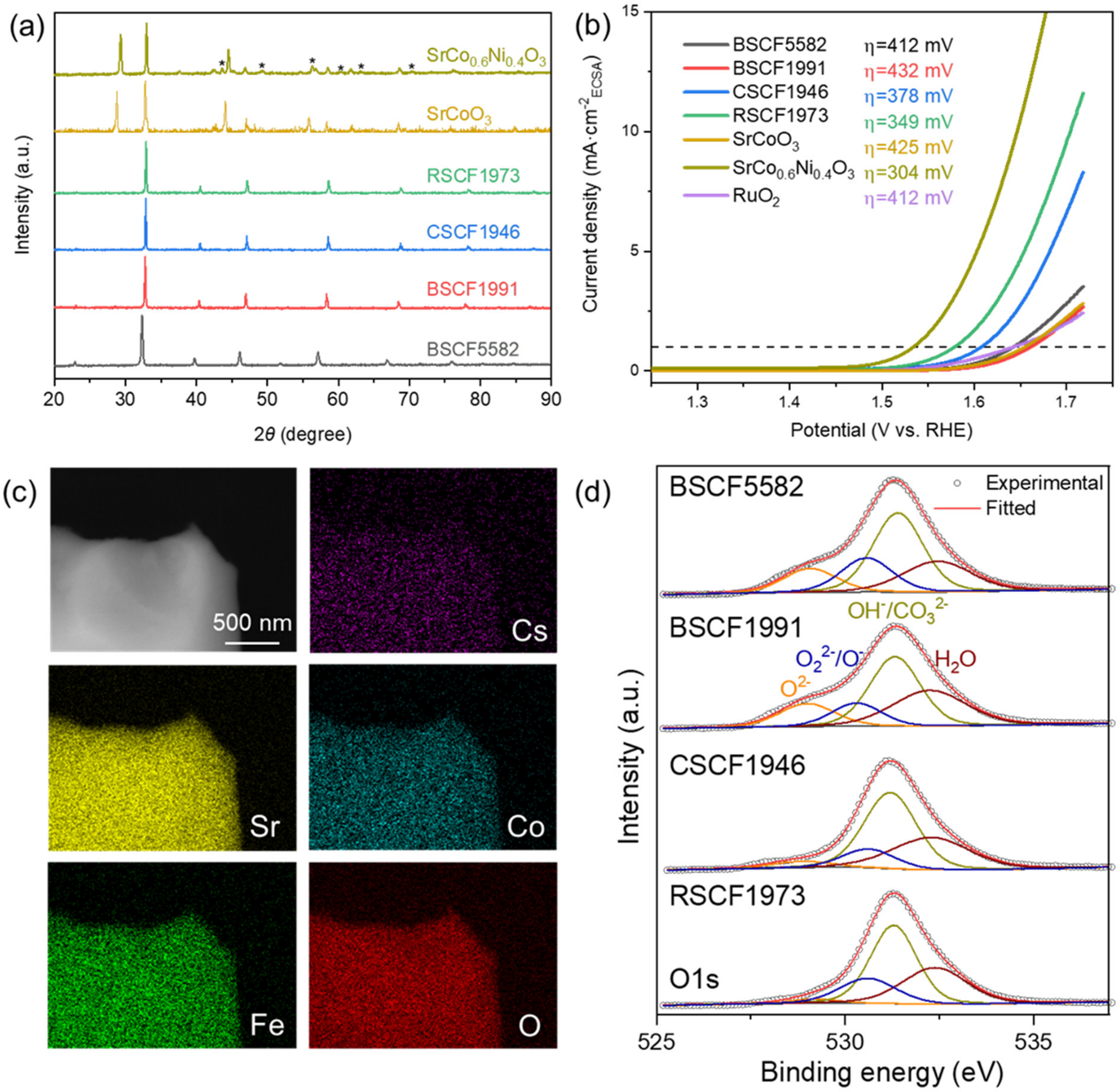

Figure 5. Experimental characterization and electrochemical performance tests of the perovskites. (a) XRD patterns. Unidentified phases are indicated with asterisks. (b) The linear sweep voltammetry (LSV) curves, with the current normalized by ECSA to evaluate the intrinsic activities[2] (See Figure S6 for ECSA measurements). Dashed line indicates the overpotential at the current densities of 1 $mA{\cdot}cm^{-2}_{ECSA}$. LSV plots normalized by the geometric area of the electrode were also provided in Figure S7(a). The overpotentials at the current densities of 10 $mA{\cdot}cm^{-2}_{geo}$ for $RuO_2$ and BSCF5582 are 341 and 405 mV, respectively, which are comparable with literature (Table S3). (c) SEM-BSE image and corresponding EDS elemental mapping of CSCF1946 (others can be found in Figure S8 and S9). (d) XPS spectra of O 1s core levels of the as-synthesized samples where four peaks were deconvoluted. The "$CO_3^{2-}$" denotes carbonate and bicarbonate species for simplicity. Each LSV curve is the average of at least two separate tests.

Further characterizations and analysis were focused on the materials that have phase-pure cubic perovskite structures: RSCF1973, CSCF1946, BSCF5582 and BSCF1991. Tafel slopes were estimated from the fitting of OER currents at different potentials sampled from steady-state chronoamperometry CA responses (Figure S7(b)-(d)), from which the four perovskites can be divided into two groups: (i) RSCF1973 & CSCF1946 and (ii) BSCF5582 & BBSCF1991. Group (i) has a Tafel slope of ~80 mV·dec$^{-1}$, and group (ii) ~60 mV·dec$^{-1}$ ($RuO_2$: 66.9 mV·dec$^{-1}$). Further increase of the applied potential lead to increased Tafel slopes in all samples, probably due to a shift of dominant reaction intermediate[28] (dashed lines in Figure S7(b)).

Morphology characterization in Figure S8 shows the perovskites are micrometer-level particles. The metal elements in the three new perovskites and the BSCF5582 are evenly distributed from the EDS elemental mapping in Figure 5(c) and Figure S9. Further, inductively coupled plasma optical emission spectroscopy (ICP-OES) shows that the synthesized perovskites have compositions very close to the nominal values except for alkali metals (Rb: 0.0031 vs. 0.1, Cs: 0.0014 vs. 0.1 (Table S4). The reduced contents for Rb and Cs are due to their volatilization during the calcination step, which can be improved by intentionally adding excessive amounts[29]. Since the predictions made by equation (2) were based on the $d_B$ and $n_B$ obtained from nominal stoichiometries, we ignored this issue and took the as-obtained phase-pure materials for further experimental examination.

XPS analysis was performed to understand the trend of intrinsic activity and OER kinetics, as shown in Figure 5(d) and Figure S10. The peaks located around 529.0 eV, 530.5 eV, 531.3 eV, and 532.4 eV are attributed to lattice oxygen $O^{2-}$, highly oxidative oxygen species $O_2^{2-}/O^{-}$[30-32], hydroxide $OH^-$/(bi)carbonate ($HCO_3^-$ and/or $CO_3^{2-}$), and adsorbed molecular water, respectively. The assignment of surface (bi)carbonate species correlates well with the observed C 1s peak at 289.2 eV which is due to the presence of (bi)carbonate[33-35] (Figure S10(c)), possibly due to the contact of samples with $CO_2$ in the atmosphere. Quantitative results of the peaks are summarized in Table S5. The oxidation states of Co and Fe in the four perovskite oxides are estimated to be around 3 and above 3, respectively, which are in line with previous reports[2,36,37]. Nevertheless, unambiguously distinguishing the oxidation states of Fe and Co is difficult from the XPS results. More details can be found in Figure S10(a)-(b).

A key message from Figure 5(d) is that the relative surface lattice oxygen concentrations on RSCF1973 and CSCF1946 are greatly less than those on both BSCFs, in line with their Tafel behaviors and suggesting the involvement of lattice oxygen in OER kinetics. The decrease of surface lattice oxygen after replacement of Ba by Rb or Cs implies increased concentration of the surface oxygen vacancy. This increased amount of vacancy of RSCF1973 and CSCF1946 over the BSCFs is understandable since they have higher $n_B$ (at nominal stoichiometries). The concentration of oxygen vacancy is related to oxygen diffusion rate and thus OER activity if LOM is involved[12]. It is worth noting that the highly oxidative oxygen species ($O_2^{2-}/O^-$) adsorbing on surface oxygen vacancy can be crucial to the OER activity[38,39], and in this work we found the OER activity correlates well with the concentration of $O_2^{2-}/O^-$ divided by the sum of $O_2^{2-}/O^-$ and $O^{2-}$ (denoted as $\delta'$ to represent the concentration of surface oxygen vacancy from XPS (Table S5). Therefore, we explain the higher activity of RSCF1973 and CSCF1946 to be caused by the increased oxygen vacancy because of the replacement of divalent Ba by monovalent Rb and Cs. In addition, we found that the logarithm of the concentration of surface oxygen vacancy can be linearly described by the descriptor ($d_B$, $n_B$), as shown in Figure S11, which again confirms the correlation between the descriptor and the surface catalytic properties.

The large presence of $OH^-/CO_3^{2-}$ may arise from the segregation of Sr on the perovskite surface[40]. In addition, substantial signals of surface Rb and Cs were observed (Figure S12), though their actual bulk compositions are below XPS detection limit 0.1 atom % (Table S4). Considering the comparable sensitivity factors of Rb and Cs with Sr, and their significantly lower bulk concentrations than Sr (e,g. Sr:Rb = ~280 in Table S4), the results reveal the enrichment of Rb and Cs on surface as well, leading to the great complexity of surface structures. Perovskite surfaces can be further complicated under realistic OER conditions because of various restructuring[26,41], which requires further studies. The data-driven descriptors and models may be refined in the future by explicitly including realistic surface information, for example, extracted from experimental spectroscopic data[42].

**In summary**, we have designed and demonstrated the efficiency of the new method SCMT-SISSO for distilling descriptors from inconsistent multi-source experimental data, which is important in the era of data-driven science. Based on 13 experimental datasets collected from different publications and the SCMT-SISSO, we obtained the 2D descriptor ($d_B$, $n_B$) for predicting the OER activity on perovskite catalysts. The descriptor ($d_B$, $n_B$) was found to be with great universality and predictive performance, and correlated with the surface properties of oxygen adsorption and oxygen-vacancy formation that explains why the OER activity increases with the $d_B$ and $n_B$ and when it may fail. This descriptor allowed us to perform high-throughput screening in a large chemical space, which yielded hundreds of unreported promising perovskites with the predicted OER activity greater than BSCF5582. Experimental validations on five candidates confirmed the three highly active new perovskite catalysts $SrCo_{0.6}Ni_{0.4}O_3$, $Rb_{0.1}Sr_{0.9}Co_{0.7}Fe_{0.3}O_3$, and $Cs_{0.1}Sr_{0.9}Co_{0.4}Fe_{0.6}O_3$.

## Methods

**SCMT-SISSO calculations**. All the 13 experimental OER activity datasets on perovskite catalysts were collected from papers published in the years from 1980 to 2020, and the Refs. are provided in the Supplementary Information. The sign-constrained multi-task learning is available in the SISSO code as of the version 3.1 at https://github.com/rouyang2017/SISSO. All the input parameters for the SCMT-SISSO calculations and the datasets can be found in the Supplementary Information.

Most of the considered perovskites in this work are solid solutions that each of the A and B-site is occupied by two different metal elements, i.e. in the general formula of $A^*_xA'_{(1-x)}B^*_yB'_{(1-y)}O_3$. The input features for machine learning were calculated from the nominal stoichiometry of the compounds, and they were defined as:

$n_B$: The total (weighted average) oxidation state of the B-site metals, calculated by $n_B = 6 - n_A$, where $n_A$ is the total oxidation state of the A-site Alkali/Alkaline-Earth/Lanthanide metal ions and can be easily obtained.

$d_B$: The average number of d-electron of the B-site metal ions, $d_B = y(D_{B^*} + 2) + (1-y)(D_{B'} + 2) - n_B$, where $D$ is the number of d-electron of a 3d metal atom.

$\chi_A$ ($\chi_B$): Pauling electronegativity for the weighted average of the A-site (B-site) metals.

$\varepsilon_B$: Bond dissociation energy of metal-oxygen diatomic molecule divided by that of oxygen-oxygen bond, as compiled in the "CRC Handbook of Chemistry and Physics", 79$^{th}$ Edition by D. R. Lide (e.d.), CRC Press, 1998.

$r_A$ ($r_B$): The averaged ionic radii of the metals at the A-site (B-site). The ionic radius is a function of the oxidation state of the ion, and thus obtaining the radii requires the input of oxidation states. $r_A$ are easy to determine since the A-site metal ions have well-defined oxidation states. Instead, the B-site metals often have variable valency, leading to the difficulty of determining the oxidation states of the $B^*$ and $B'$, and hence their ionic radii. For this, we designed the following scheme to calculate the oxidation states and ionic radii for the $B^*$ and B':

(1) Set the intervals for the possible oxidation states of each species according to its common oxidation states, and let the oxidation state $n_{B^*}$ and $n_{B'}$ to vary at the step of 0.1. The intervals used in this work were [1,4] for Ti, [1, 5] for V, Cr, Mn, Fe, Co, [1, 4] for Ni, and [1, 3] for Cu.

(2) Take the values for $n_{B^*}$ and $n_{B'}$ such that $n_B = y \cdot n_{B^*} + (1-y) \cdot n_{B'}$ and the ionic electronegativity of B equal that of B', with $\chi = \chi_{AR} + 0.359q/r^2$, where $\chi_{AR}$ is the Allred-Rochow electronegativity of atom[43]. The second term on the right side is the correction of electrostatic force due to the charge. The Shannon ionic radii $r$ for any oxidation states were taken from Ref.[44]

$\mu, t$: The octahedral factor and tolerance factor of the perovskite structure, respectively.

**DFT calculations**. Spin-polarized DFT calculations were performed by using the Vienna Ab initio Simulation Package (VASP)[45], with the Perdew−Burke−Ernzerhof (PBE) functional under the generalized gradient approximation[46] and the projected augmented wave (PAW) pseudopotentials[47]. The GGA+U approach by Dudarev et al.[48] was employed to treat the on-site Coulomb interactions on the localized 3d electrons, with the U values taken from the Ref.[49]. The energy cutoff of 450 eV was used for the plane-wave basis set in all the calculations.

In the calculations of O adsorption and vacancy formation in Figure 3(a), the B-site terminated (100) surfaces of single perovskites in cubic structures were modeled by using $2 \times 2$ slabs with six atomic layers. The k-points mesh of $5 \times 5 \times 1$ were used to sample the Brillouin zone. All atoms were allowed to fully relax until the residual forces were less than 0.03 eV/Å, except for the bottom three layers that were fixed to their bulk positions. For the calculations of bulk BSCF5582 and RSCF2846 in Figure S4, the supercells of $4 \times 5 \times 5$ in cubic structures were used, with their atomic configurations generated by using the Special Quasirandom Structure (SQS)[27] method. Only the gamma-point for the k-points mesh was used for these big supercells (500 atoms), and all atoms were allowed to fully relax. The optimized lattice constants with the SQS models was 3.99 Å per $ABX_3$ formula for the BSCF (equal to the experimental value: 3.99 Å)[2], and was 3.93 Å for the RSCF.

**Materials and synthesis**. The perovskite oxides listed in Figure 5 and Figure S5 were prepared by using a sol-gel method[50]. Certain amounts of metal nitrate (or acetate or metavanadate) according to the nominal ratio were dissolved in 200 mL ultra-pure Milli-Q water (18 MΩ·cm) at 80 ℃. $Ba(NO_3)_2$ (>99.5%), $Sr(NO_3)_2$ (>99.0%), and $Ni(NO_3)_2 \cdot 6H_2O$ (>98.0%)were purchased from Sinopharm Chemical Reagent Co., Ltd. $Co(NO_3)_2 \cdot 6H_2O$ (99.99%), $Fe(NO_3)_3 \cdot 9H_2O$ (analytical reagent), $Cu(NO_3)_2 \cdot 3H_2O$ (99.0%), $Mn(NO_3)_2 \cdot 4H_2O$ (98%), lanthanum acetate hydrate (99.9%) and $NH_4VO_3$ (ACS reagent) were purchased from Aladdin. $RbNO_3$ (99.0%) was purchased from Macklin and $CsNO_3$ (>99%) from Adamas-beta. $Ca(NO_3)_2 \cdot 4H_2O$ (>98.0%) was purchased from Shanghai Titan Scientific Co., Ltd. All chemicals were used without further purification. Then ethylenedinitrilotetraacetic acid (EDTA, ACS reagent from Sigma-Aldrich) and citric acid (CA, 99.8% from Aladdin and 99.5% from Energy chemical) as chelating agents were added with stirring for 1 h. The pH was then adjusted to around 8 by adding ammonia solution (guaranteed reagent from Sinopharm). The solution was evaporated to form a gel under continuous stirring in an oil bath set at 120 ℃, which was later transferred to oven at 250 ℃ for 8 h. The formed black powders were finally calcined in muffle furnace at 900 ℃ for 5 h at a ramping rate of 10 ℃/min.

**Characterizations**. XRD patterns of perovskite oxides were measured with a X-ray diffractometer (Bruker D2 PHASER) using Cu Kα radiation (λ= 1.5418 Å). The morphology of the perovskite particles was investigated using a scanning electron microscope (SEM, Zeiss GeminiSEM 300) operated at 15 kV. Elemental mapping was carried out by an energy dispersive X-ray spectrometer integrated on the SEM (Oxford Instruments). X-ray photoelectron spectroscopy analysis was performed using a Thermo Scientific ESCALAB 250Xi equipped with an Al Kα radiation source (1486.6 eV). Details of the X-ray photoelectron spectroscopy fittings are given in Table S5. ICP-OES was carried out on an Agilent 730 spectrometer to analyze the actual composition of synthesized perovskite oxides.

**Electrochemical measurements**. Catalyst ink was prepared by sonicating a mixture of 5 mg perovskite material, 0.5 mL water, 0.5 mL ethanol (>99.7%, Sinopharm) and 0.01 mL Nafion solution (5wt% from Sigma-Aldrich) for 60 min. 0.01 mL sonicated mixture was then drop casted on a glassy carbon electrode with an area of 0.196 $cm^2$ followed by air-drying for 30 min. The catalyst loading on the electrode is 0.25 $mg/cm^2$. The electrochemical measurements were performed in a three-electrode configuration controlled by Autolab potentiostat (Metrohm). The glassy carbon electrode with catalyst was used as a working electrode and was rotated using a rotating disk electrode system at 1600 rpm. The Hg/HgO electrode was used as a reference electrode and graphite electrode as counter electrode. The electrolyte of 1 M KOH (pH=13.9) was prepared by using KOH pellets (99.99%, Aladdin) and Milli-Q water. The electrolyte was purged by $O_2$ (99.999%) for 20 min to remove any impurities and maintain an $O_2$-saturated condition. The potential values versus Hg/HgO were converted to be expressed against RHE according to the equation

$$E_{RHE} = E_{Hg/HgO} + 0.059 \text{ pH} + 0.098$$

The catalyst was initially conditioned by performing CV cycling with a scan rate of 100 mV/s between 0.9 V and 1.7 V vs. RHE for 15-30 cycles until the performance is stable. Then the LSV tests with a slow scan rate of 5 mV/s were performed to evaluate the activity of perovskite oxides. Ohmic drop was corrected (95% compensation) for all the polarization curves where the electrolyte resistance was determined by electrochemical impedance spectroscopy. Electrochemically active surface area (ECSA) was determined by CV (SCAN250 module at Autolab) at various scan rates (10 mV/s – 120 mV/s) recorded at open circuit potential where there is no faradaic current response. Please note that the electrolyte was purged by $N_2$ (99.999%) prior to the ECSA tests. The double layer capacitive current was taken as the average of the anodic and cathodic currents. The Tafel plot was constructed by the steady-state currents collected from multistep CA in a specific potential range determined from LSV (linear region in E vs. log($i$) curves). Each measurement was conducted at least two times and the results were represented as the averaged ones. $RuO_2$ (99.9%, Sigma-Aldrich) was used as a reference for comparison.

## Acknowledgements

This project is supported by the National Natural Science Foundation of China (Grant No. 22173058). R.O. acknowledges the Key Research Project of Zhejiang Laboratory (Grant No. 2021PE0AC02). Y. W. would like to sincerely thank Prof. Feng at Shanghai University for her kind supports for some of the experimental apparatus in our experiments.

## Author contributions

R.O. conceived this project. R.O. supervised the machine learning and DFT calculations. Y.W. supervised the experiments. J.W. performed the machine learning and DFT calculations. H.X. performed the experiments. R.O. and Y.W. wrote the manuscript with inputs from all the authors. All authors contributed to the analysis and interpretation of the results, and commented on the manuscript.

## Competing interests

The authors declare no competing interests.

## References

1 Hong, W. T., Risch, M., Stoerzinger, K. A., Grimaud, A., Suntivich, J. & Shao-Horn, Y. Toward the rational design of non-precious transition metal oxides for oxygen electrocatalysis. *Energy Environ. Sci.* **8**, 1404-1427 (2015).

2 Suntivich, J., May, K. J., Gasteiger, H. A., Goodenough, J. B. & Shao-Horn, Y. A Perovskite oxide optimized for oxygen evolution catalysis from molecular orbital principles. *Science* **334**, 1383-1385 (2011).

3 An, L., Wei, C., Lu, M., Liu, H., Chen, Y., Scherer, G. G., Fisher, A. C., Xi, P., Xu, Z. J. & Yan, C.-H. Recent development of oxygen evolution electrocatalysts in acidic environment. *Adv. Mater.* **33**, 2006328 (2021).

4 Rossmeisl, J., Qu, Z.-W., Zhu, H., Kroes, G.-J. & Nørskov, J. K. Electrolysis of water on oxide surfaces. *J. Electroanal. Chem.* **607**, 83-89 (2007).

5 Pan, Y., Xu, X., Zhong, Y., Ge, L., Chen, Y., Veder, J.-P. M., Guan, D., OHayre, R., Li, M., Wang, G., Wang, H., Zhou, W. & Shao, Z. Direct evidence of boosted oxygen evolution over perovskite by enhanced lattice oxygen participation. *Nat. Commun.* **11**, 2002 (2020).

6 Grimaud, A., Diaz-Morales, O., Han, B., Hong, W. T., Lee, Y.-L., Giordano, L., Stoerzinger, K. A., Koper, M. T. M. & Shao-Horn, Y. Activating lattice oxygen redox reactions in metal oxides to catalyse oxygen evolution. *Nat. Chem.* **9**, 457-465 (2017).

7 Hong, W. T., Stoerzinger, K. A., Lee, Y.-L., Giordano, L., Grimaud, A., Johnson, A. M., Hwang, J., Crumlin, E. J., Yang, W. & Shao-Horn, Y. Charge-transfer-energy-dependent oxygen evolution reaction mechanisms for perovskite oxides. *Energy Environ. Sci.* **10**, 2190-2200 (2017).

8 Liu, J., Liu, H., Chen, H., Du, X., Zhang, B., Hong, Z., Sun, S. & Wang, W. Progress and challenges toward the rational design of oxygen electrocatalysts based on a descriptor approach. *Adv. Sci.* **7**, 1901614 (2020).

9 Bockris, J. O. M. & Otagawa, T. The electrocatalysis of oxygen evolution on perovskites. *J. Electrochem. Soc.* **131**, 290-302 (1984).

10 Man, I. C., Su, H.-Y., Calle-Vallejo, F., Hansen, H. A., Martinez, J. I., Inoglu, N. G., Kitchin, J., Jaramillo, T. F., Nørskov, J. K. & Rossmeisl, J. Universality in oxygen evolution electrocatalysis on oxide surfaces. *ChemCatChem* **3**, 1159-1165 (2011).

11 Grimaud, A., May, K. J., Carlton, C. E., Lee, Y.-L., Risch, M., Hong, W. T., Zhou, J. & Shao-Horn, Y. Double perovskites as a family of highly active catalysts for oxygen evolution in alkaline solution. *Nat. Commun.* **4**, 2439 (2013).

12 Mefford, J. T., Rong, X., Abakumov, A. M., Hardin, W. G., Dai, S., Kolpak, A. M., Johnston, K. P. & Stevenson, K. J. Water electrolysis on $La_{1-x}Sr_xCoO_{3-\delta}$ perovskite electrocatalysts. *Nat. Commun.* **7**, 11053 (2016).

13 Weng, B., Song, Z., Zhu, R., Yan, Q., Sun, Q., Grice, C. G., Yan, Y. & Yin, W.-J. Simple descriptor derived from symbolic regression accelerating the discovery of new perovskite catalysts. *Nat. Commun.* **11**, 3513 (2020).

14 Yamada, I., Takamatsu, A., Asai, K., Shirakawa, T., Ohzuku, H., Seno, A., Uchimura, T., Fujii, H., Kawaguchi, S., Wada, K., Ikeno, H. & Yagi, S. Systematic study of descriptors for oxygen evolution reaction catalysis in perovskite oxides. *J. Phys. Chem. C* **122**, 27885-27892 (2018).

15 Hong, W. T., Welsch, R. E. & Shao-Horn, Y. Descriptors of oxygen-evolution activity for oxides: A statistical evaluation. *J. Phys. Chem. C* **120**, 78-86 (2016).

16 Caruana, R. Multitask learning. *Mach. Learn.* **28**, 41-75 (1997).

17 Obozinski, G., Taskar, B. & Jordan, M. Multi-task feature selection. *Tech. Rep. Department of Statistics, University of California, Berkeley*, (2006).

18 Ouyang, R., Ahmetcik, E., Carbogno, C., Scheffler, M. & Ghiringhelli, L. M. Simultaneous learning of several materials properties from incomplete databases with multi-task SISSO. *J. Phys.: Mater.* **2**, 024002 (2019).

19 Ouyang, R., Curtarolo, S., Ahmetcik, E., Scheffler, M. & Ghiringhelli, L. M. SISSO: A compressed-sensing method for identifying the best low-dimensional descriptor in an immensity of offered candidates. *Phys. Rev. Mater.* **2**, 083802 (2018).

20 Friedman, J., Hastie, T., Hofling, H. & Tibshirani, R. Pathwise coordinate optimization. *Ann. Appl. Stat.* **1**, 302-332 (2007).

21 Calle-Vallejo, F., Diaz-Morales, O. A., Kolb, M. J. & Koper, M. T. M. Why is bulk thermochemistry a good descriptor for the electrocatalytic activity of transition metal oxides? *ACS Catal.* **5**, 869-873 (2015).

22 Calle-Vallejo, F., Inoglu, N. G., Su, H.-Y., Martınez, J. I., Man, I. C., Koper, M. T. M., Kitchin, J. R. & Rossmeisl, J. Number of outer electrons as descriptor for adsorption processes on transition metals and their oxides. *Chem. Sci.* **4**, 1245 (2013).

23 Hammer, B. & Nørskov, J. K. Why gold is the noblest of all the metals. *Nature* **376**, 238-240 (1995).

24 Wexler, R. B., Gautam, G. S., Stechel, E. B. & Carter, E. A. Factors governing oxygen vacancy formation in oxide perovskites. *J. Am. Chem. Soc.* **143**, 13212-13227 (2021).

25 Bartel, C. J., Sutton, C., Goldsmith, B. R., Ouyang, R., Musgrave, C. B., Ghiringheli, L. M. & Scheffler, M. New tolerance factor to predict the stability of perovskite oxides and halides. *Sci. Adv.* **5**, eaav0693 (2019).

26 May, K. J., Carlton, C. E., Stoerzinger, K. A., Risch, M., Suntivich, J., Lee, Y. L., Grimaud, A. & Shao-Horn, Y. Influence of oxygen evolution during water oxidation on the surface of perovskite oxide catalysts. *J. Phys. Chem. Lett.* **3**, 3264-3270 (2012).

27 Zunger, A., Wei, S.-H., Ferreira, L. G. & Bernard, J. E. Special quasirandom structures. *Phys. Rev. Lett.* **65**, 353-356 (1990).

28 Shinagawa, T., Garcia-Esparza, A. T. & Takanabe, K. Insight on Tafel slopes from a microkinetic analysis of aqueous electrocatalysis for energy conversion. *Sci. Rep.* **5**, 13801 (2015).

29 Yoshimura, J., Ebina, Y., Kondo, J., Domen, K. & Tanaka, A. Visible light-induced photocatalytic behavior of a layered perovskite-type rubidium lead niobate, $RbPb_2Nb_3O_{10}$. *J. Phys. Chem.* **97**, 1970-1973 (1993).

30 Merino, N. A., Barbero, B. P., Eloy, P. & Cadús, L. E. $La_{1-x}Ca_xCoO_3$ perovskite-type oxides: Identification of the surface oxygen species by XPS. *Appl. Surf. Sci.* **253**, 1489-1493 (2006).

31 Wang, Y., Ren, J., Wang, Y., Zhang, F., Liu, X., Guo, Y. & Lu, G. Nanocasted synthesis of mesoporous $LaCoO_3$ perovskite with extremely high surface area and excellent activity in methane combustion. *J. Phys. Chem. C* **112**, 15293-15298 (2008).

32 Xu, X., Chen, Y., Zhou, W., Zhu, Z., Su, C., Liu, M. & Shao, Z. A perovskite electrocatalyst for efficient hydrogen evolution reaction. *Adv. Mater.* **28**, 6442-6448 (2016).

33 Stoch, J. & Gablankowska-Kukucz, J. The effect of carbonate contaminations on the XPS O 1s band structure in metal oxides. *Surf. Interface Anal.* **17**, 165-167 (1991).

34 Gonzalez-Elipe, A. R., Espinos, J. P., Fernandez, A. & Munuera, G. XPS study of the surface carbonation/hydroxylation state of metal oxides. *Appl. Surf. Sci.* **45**, 103-108 (1990).

35 Stoerzinger, K. A., Hong, W. T., Crumlin, E. J., Bluhm, H., Biegalski, M. D. & Shao-Horn, Y. Water reactivity on the $LaCoO_3$(001) surface: An ambient pressure X-ray photoelectron spectroscopy study. *J. Phys. Chem.*

*C* **118**, 19733-19741 (2014).

36 Arnold, M., Xu, Q., Tichelaar, F. D. & Feldhoff, A. Local charge disproportion in a high-performance perovskite. *Chem. Mater.* **21**, 635-640 (2009).

37 Harvey, A. S., Litterst, F. J., Yang, Z., Rupp, J. L. M., Infortuna, A. & Gauckler, L. J. Oxidation states of Co and Fe in $Ba_{1-x}Sr_xCo_{1-y}Fe_yO_{3-\delta}$ (x, y = 0.2–0.8) and oxygen desorption in the temperature range 300–1273 K. *Phys. Chem. Chem. Phys.* **11**, 3090-3098 (2009).

38 Zhu, J., Li, H., Zhong, L., Xiao, P., Xu, X., Yang, X., Zhao, Z. & Li, J. Perovskite oxides: preparation, characterizations, and applications in heterogeneous catalysis. *ACS Catal.* **4**, 2917-2940 (2014).

39 Zhu, Y., Zhou, W., Yu, J., Chen, Y., Liu, M. & Shao, Z. Enhancing electrocatalytic activity of perovskite oxides by tuning cation deficiency for oxygen reduction and evolution reactions. *Chem. Mater.* **28**, 1691-1697 (2016).

40 Koo, B., Kim, K., Kim, J. K., Kwon, H., Han, J. W. & Jung, W. Sr segregation in perovskite oxides: Why it happens and how it exists. *Joule* **2**, 1476-1499 (2018).

41 Chen, S., Ma, L., Huang, Z., Liang, G. & Zhi, C. In situ/operando analysis of surface reconstruction of transition metal-based oxygen evolution electrocatalysts. *Cell Rep. Phys. Sci.* **3**, 100729 (2022).

42 Timoshenko, J. Spectroscopy predicts catalyst functionality. *Nat. Catal.* **5**, 469-470 (2022).

43 Allred, A. L. & Rochow, E. G. A scale of electronegativity based on electrostatic force. *J. Inorg. Nucl. Chem.* **5**, 264 (1958).

44 Ouyang, R. Exploiting ionic radii for rational design of halide perovskites. *Chem. Mater.* **32**, 595-604 (2020).

45 Kresse, G. & Furthmuller, J. Efficient iterative schemes for ab initio total-energy calculations using a plane-wave basis set. *Phys. Rev. B* **54**, 11169 (1996).

46 Perdew, J. P., Burke, K. & Ernzerhof, M. Generalized gradient approximation made simple. *Phys. Rev. Lett.* **77**, 3865 (1996).

47 Blöchl, P. E. Projector augmented-wave method. *Phys. Rev. B* **50**, 17953 (1994).

48 Dudarev, S. L., Botton, G. A., Savrasov, S. Y., Humphreys, C. J. & Sutton, A. P. Electron-energy-loss spectra and the structural stability of nickel oxide: An LSDA+U study. *Phys. Rev. B* **57**, 1505 (1998).

49 Wang, L., Maxisch, T. & Ceder, G. Oxidation energies of transition metal oxides within the GGA+U framework. *Phys. Rev. B* **73**, 195107 (2006).

50 Zhou, W., Zhao, M., Liang, F., Smith, S. C. & Zhu, Z. High activity and durability of novel perovskite electrocatalysts for water oxidation. *Mater. Horiz.* **2**, 495-501 (2015).